\documentclass[aps,prx,showpacs,twocolumn,groupedaddress,superscriptaddress]{revtex4-1}

\usepackage{times,xspace}
\usepackage{graphicx,graphics,color,epsfig}
\usepackage{amsmath}
\usepackage{amssymb}
\usepackage{amsfonts}
\usepackage{amsfonts}
\usepackage{epstopdf}
\usepackage{bm}
\usepackage{hyperref}
\usepackage{color}
\usepackage{float}
\usepackage[normalem]{ulem}
\usepackage{comment}

\def\be{\begin{equation}}
\def\ee{\end{equation}}
\def\ba{\begin{eqnarray}}
\def\ea{\end{eqnarray}}

\begin{document}

\title{Photo-induced Ferromagnetic and Superconducting Orders in Multi-orbital Hubbard Models}

\author{Sujay Ray}
\author{Philipp Werner}
\affiliation{Department of Physics, University of Fribourg, 1700 Fribourg, Switzerland}

\date{\today}
 
\begin{abstract}
The search for hidden orders in photoexcited lattice systems is an active research field driven by experimental reports of light-induced or light-stabilized phases. In this study, we investigate hidden electronic orders in strongly correlated two-orbital Hubbard models with orbital-dependent bandwidths. In equilibrium, the half-filled systems are antiferromagnetically ordered. Using non-equilibrium dynamical mean field theory we demonstrate the appearance of nonthermal ferromagnetic order in the photo-doped state, if the two bandwidths are sufficiently different, and its coexistence with spin-singlet $\eta$-superconductivity in the high photo-doping region. Spin-triplet $\eta$-superconducting order appears instead if the two bandwidths are comparable. The rich nonequilibrium phasediagram uncovered in this work shows that Mott insulating multi-orbital systems provide an interesting platform for the realization of nonthermal electronic orders.  
\end{abstract}
\vspace{0.5in}

\maketitle

{\it{Introduction}}:  The exploration of metastable or hidden phases, which cannot be accessed under equilibrium conditions, is an exciting research direction in condensed matter physics \cite{lat_struc, mang_hid,Fausti2011,Cavalleri2018,Torre2021,ph_noneq}. Due to the interplay between different active degrees of freedom \cite{Dagotto2005}, strongly correlated electron systems are particularly interesting candidates for the realization of such hidden phases \cite{yuta_etasc,yuta_etasc1,opt_magord, ind_sc_cuprate, phind_sc}. Among them, Mott insulators have been shown to host different types of hidden phases \cite{opt_exint, eta_jiajun, eta_spintrip, tri_chiral, exciton_insulator, Kugel_Khomskii} when excited by a laser pulse. Recent theoretical studies of photo-doped Mott insulators identified nonthermal staggered superconducting (SC) orders in single-band and two-band Hubbard models \cite{yuta_etasc, yuta_etasc1, kaneko1, kaneko2, eta_jiajun, eta_spintrip} and chiral superconducting orders in geometrically frustrated systems \cite{tri_chiral}. Two-orbital Hubbard models have also been shown to exhibit nonthermal excitonic order \cite{exciton_insulator} and Kugel-Khomoskii spin-orbital order \cite{Kugel_Khomskii}. Because of the long lifetime of charge excitations in large-gap Mott insulators \cite{Sensarma2010,Eckstein2011,Strohmaier2010,Lenarcic2014}, a partial thermalization of charge carriers may occur in each Hubbard band, and these photo-doped metastable states can be susceptible to new types of ordering instabilities. Due to the rapid advancement in laser technology, it becomes possible to induce and detect such metastable states \cite{VO2_dynamics,ferro_chrgtrnsfr_dynamics}, which also provides interesting perspectives for technological applications.

Several experimental and theoretical studies considered magnetic orders in photoexcited systems \cite{mag_switch,opt_magord, manganites_1,VO2_dynamics,ferro_chrgtrnsfr_dynamics, Qint_Mott,Ishihara_2007,Ishihara_2011,Ishihara_DE1,Ishihara_DE2,Qdyn_spin_struc, opt_magcont}. The phenomena range from ultrafast demagnetization \cite{ultrafast_demag1, ultrafast_demag2} to magnetic switching and the appearance of nonthermal magnetic orders \cite{mag_switch, Kugel_Khomskii}. The double exchange model plays a prominent role in studies of photo-induced magnetic phases. In the context of perovskite manganites, a series of numerical investigations based on exact diagonalization and density matrix renormalization group calculations reported nonthermal magnetic phases \cite{manganites_1,manganites_2, manganites_3, Ishihara_DE1, Ishihara_DE2, Ishihara_DE3}. However, these studies of closed systems lack a systematic control over the density and effective temperature of the photo-carriers. While the nonthermal melting of Kugel-Khomskii spin-orbital order in a photo-doped system has been investigated with non-equilibrium dynamical mean field theory (DMFT) \cite{Kugel_Khomskii}, the emergence of magnetic order is difficult to study in real-time simulations, due to heating effects and intrinsically slow dynamics. 

Here we use the Nonequilibrium Steady State (NESS) approach \cite{NESS_martin}, which provides a high level of control over the effective temperature and density of photo-carriers, to study the two-orbital Hubbard model with orbital-dependent bandwidths. The interplay between more localized and more itinerant electrons leads to nontrivial couplings between the spin, orbital, and charge degrees of freedom and results in a multitude of hidden electronic orders.

\begin{figure}[t]
\includegraphics[width=0.4\textwidth]{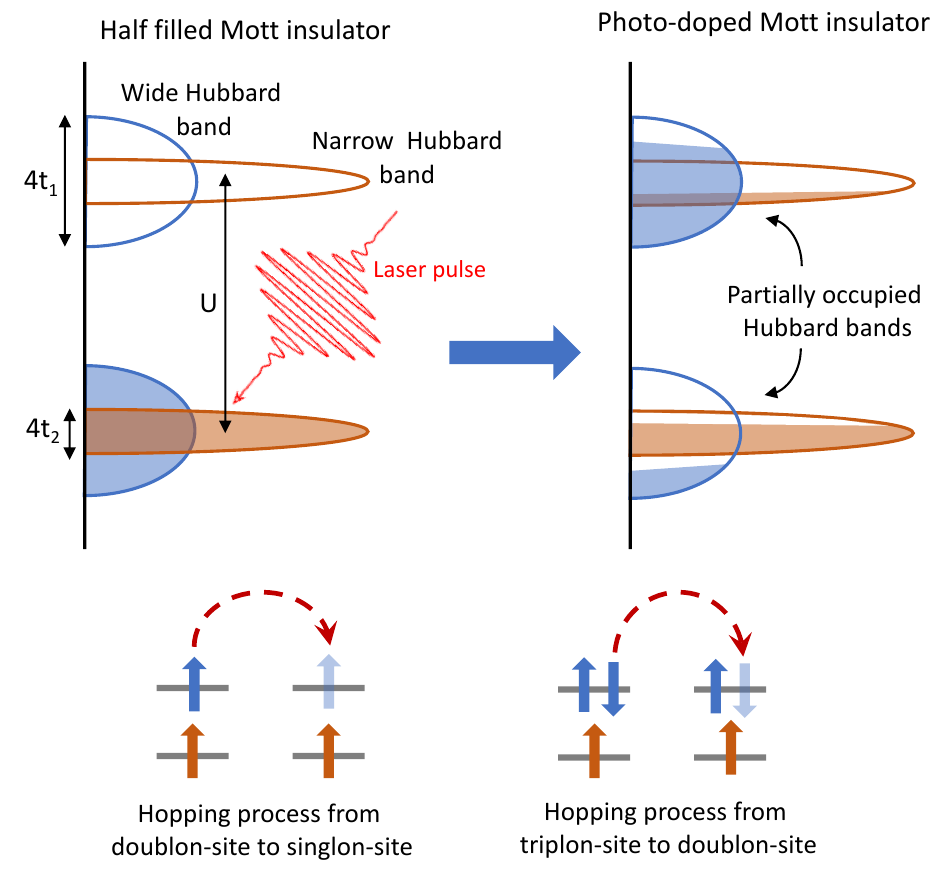}
\caption{ Schematic diagram showing the photo-excitation in two-orbital Mott insulators with different bandwidths. After the applied laser pulse, the wider band is photo-doped more strongly than the narrower band. $t^\text{hop}_1$ and $t^\text{hop}_2$ are the hopping parameters for orbital $1$ and $2$, respectively. The bottom panels show the dominant hopping processes at holon/triplon density 0.25, which stabilize the FM state.
}
\label{fig1}
\end{figure}

{\it{Model and Method}}:  We consider the two-orbital Hubbard model with Hamiltonian
\begin{align}
&H = -\sum_{\left<ij\right>,\sigma} \sum_{\alpha=1,2} t^\text{hop}_{\alpha}c^{\dagger}_{i,\alpha\sigma}c_{j,\alpha\sigma} \nonumber \\
&\hspace{0mm} + U\sum_{i}\sum_{\alpha=1,2}n_{i,\alpha\uparrow}n_{i,\alpha\downarrow} - \mu \sum_{i}\sum_{\alpha=1,2}\left(n_{i,\alpha\uparrow} + n_{i,\alpha\downarrow} \right) \nonumber \\
&\hspace{0mm} + (U-2J)\sum_{i,\sigma}n_{i,1\sigma}n_{i,2\Bar{\sigma}} 
+ (U-3J)\sum_{i,\sigma}n_{i,1\sigma}n_{i,2\sigma} ,\label{eq_1}
\end{align}
where $c^{\dagger}_{i,\alpha\sigma}$ $(c_{i,\alpha\sigma} \nonumber)$ is the creation (annihilation) operator at site $i$, $\alpha$ and $\sigma$ denote the orbital and spin indices, respectively, $t^\text{hop}_{\alpha}$ is the orbital dependent nearest neighbor hopping amplitude between sites $i$ and $j$, $U$ is the intra-orbital Hubbard repulsion, $J$ the Hund coupling, and $\mu$ the chemical potential. In general, we assume different hopping parameters for the two orbitals and hence different bandwidths. We are interested in the strongly correlated regime of the half filled Hubbard model, where both bands are Mott insulating, see the upper panel of Fig.~\ref{fig1}. When a laser pulse is applied to such a system, the wider band gets more strongly photo-doped.

In the NESS simulations \cite{NESS_martin}, the system is weakly coupled to cold Fermion baths at each site. These baths induce charge carriers (triply and singly occupied sites) into our half-filled system, thus creating an effective photoexcited state. The Fermion baths are aligned with the lower and the upper Hubbard bands. By choosing the chemical potential of the Fermion baths $\mu_b$, we can control the density of photo-carriers, while changing the temperature of the Fermion baths $T_b$ allows to control the effective temperature of the photo-doped system. We can thus map out the nonequilibrium phase diagram of the photo-excited system as a function of photo-doping concentration and effective temperature.

The DMFT simulations of the NESS are implemented for an infinitely connected Bethe lattice using the non-crossing approximation (NCA) as impurity solver \cite{Keiter_1970, Eckstein_2010}. The DMFT self-consistency equation in the Nambu Keldysh formalism takes the form
\begin{eqnarray}
    \Delta(t,t^{\prime})=t_{0}^{2}\gamma G(t,t^{\prime})\gamma + \sum_{b} D_{b}(t,t^{\prime}).
\label{eq_2}
\end{eqnarray}
Here $\Delta$ is the hybridization function, which has two contributions. The first term represents the lattice self-consistency for the Bethe lattice, with $t_{0}=W/2$, where $W$ is the half-bandwidth. The second term comes from the Fermionic baths and is defined in frequency space by $D_b(\omega)=$ $g^2\rho_{b}(\omega)=\Gamma \sqrt{W_{b}^{2}-(\omega - \omega_{b})^{2}}$, where $\Gamma=g^2/W_{b}^{2}$ is a dimensionless coupling constant, $\omega_b$ indicates the center of the energy spectrum and $W_b$ indicates the half-bandwidth of the bath $b$. In Eq.~\eqref{eq_2} we use the spinor basis $\psi^{\dagger}=(c^{\dagger}_{1\uparrow} \hspace{0.15cm} c^{\dagger}_{2\uparrow} \hspace{0.15cm} c_{1\downarrow} \hspace{0.15cm} c_{2\downarrow})$ to study spin-singlet SC order and $\psi^{\dagger}=(c^{\dagger}_{1\uparrow} \hspace{0.15cm} c_{2\uparrow} \hspace{0.15cm} c^{\dagger}_{1\downarrow} \hspace{0.15cm} c_{2\downarrow})$ for spin-triplet SC order. The choice of the spinor $\psi^{\dagger}$ and the $\gamma$ matrix allows us to study different magnetic and SC phases, see Supplemental Material (SM). Because of the steady state assumption, the hybridization functions and Green's functions depend only on the time difference $t-t^{\prime}$.

In all our calculations we use four fermion baths with $\omega_{b}=\pm U/2, \pm 3U/2$ and  $W_b=3.0$ and $\Gamma=0.028$ for efficient photo-doping and we use the interaction parameters $U=20$ and $J=1$. The hopping parameter for orbital $1$ is kept fixed at $t^\text{hop}_{1}=1$, while $t^\text{hop}_{2}$ for orbital $2$ is varied from $0.1$ to $1.0$. We measure the photo-doping concentration by the triplon density, which we define as $\left< n_{1\uparrow}n_{1\downarrow}n_{2}+n_{1}n_{2\uparrow}n_{2\downarrow}-4n_{1\uparrow}n_{1\downarrow}n_{2\uparrow}n_{2\downarrow} \right>$, with $n_{\alpha}=n_{\alpha\uparrow} + n_{\alpha\downarrow}$ the occupation of orbital $\alpha$. In addition, we define the magnetic orders via the local magnetization $m_{i}=\sum_{\alpha} n_{i,\alpha\uparrow} - n_{i,\alpha\downarrow}$, and two types of SC order - spin-singlet orbital-triplet order $p_{i,oy}=\frac{1}{4}\sum_{\alpha} ( c^{\dagger}_{i,\alpha\uparrow} c^{\dagger}_{i,\alpha\downarrow} + h.c. )$, and spin-triplet orbital-singlet SC order $p_{i,sy}=\frac{1}{4}\sum_{\sigma} ( c^{\dagger}_{i,1\sigma} c^{\dagger}_{i,2\sigma} + h.c. )$. In a staggered SC state ($\eta$ order) or staggered antiferromagnetic state (AFM order) these order parameters have opposite signs on the two sublattices $A$ and $B$, 
\begin{eqnarray}
\eta^{x}_{i,oy(sy)}&=&\delta_i^{A/B}p_{i,oy(sy)}, \\
M_{i,AFM}&=&\delta_i^{A/B}m_{i},
\label{eq_5}
\end{eqnarray}
with $\delta_i^{A(B)}=1$ $(-1)$ for $i$ on the $A$ $(B)$ sublattice.

In order to study different phases we apply small seed fields which couple to the order parameters of interest. In a symmetry-broken state the order parameter grows to a high value and becomes almost independent of the value of the seed field. In the DMFT calculations, we switch off the seed field after some iterations, so that the symmetry-broken phase can be unambiguously identified by a non-zero order-parameter. We cannot simultaneously realize $\eta^{x}_{i,oy}$ and $\eta^{x}_{i,sy}$ SC orders, since they correspond to different $\psi^{\dagger}$ and $\gamma$, while FM and $\eta^{x}_{i,oy}$ order can be simultaneously stabilized.

\begin{figure}[t]
\includegraphics[width=0.48\textwidth]{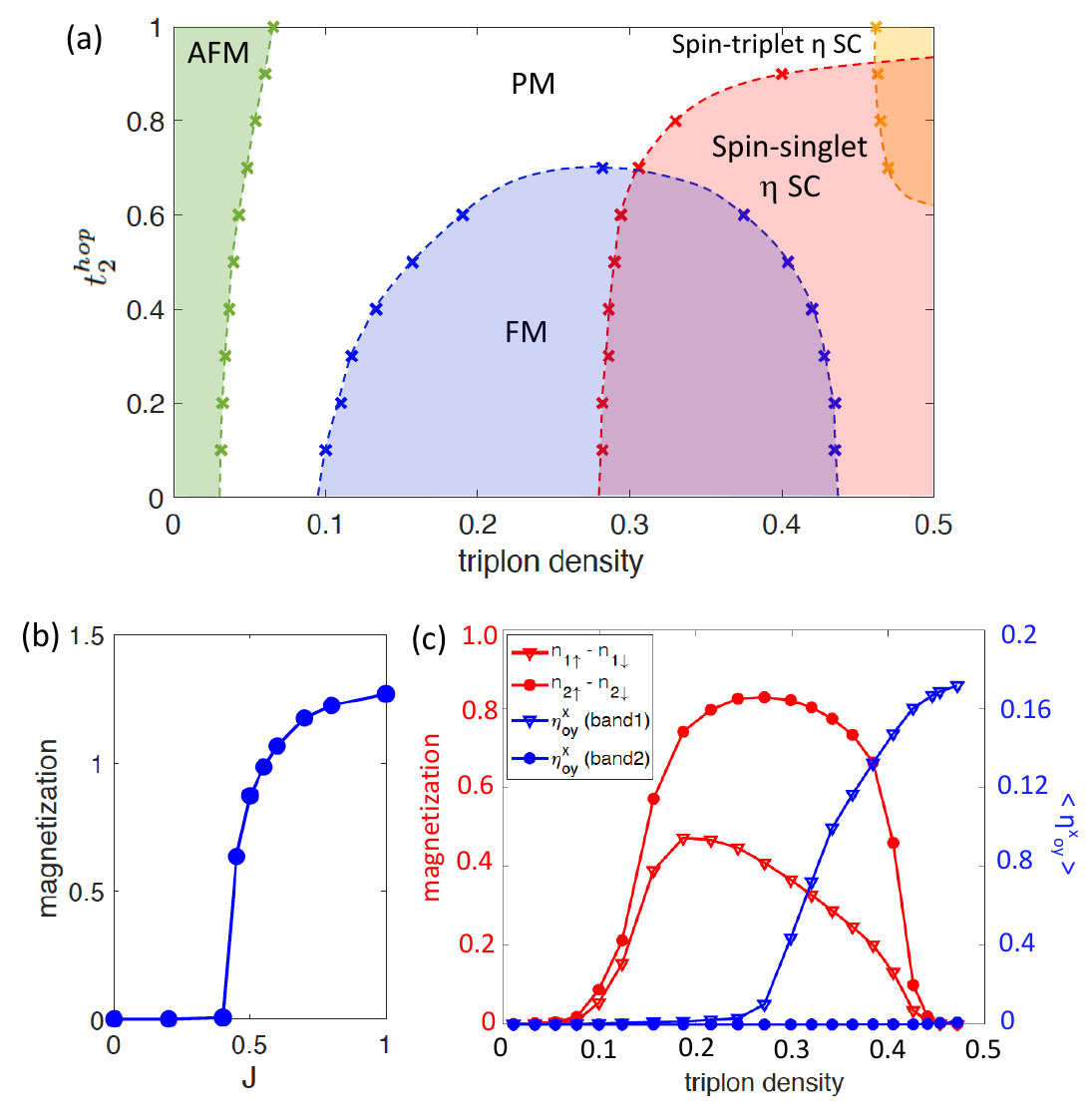}
\caption{(a) Nonequilibrium phase diagram of the photo-doped half-filled two-orbital Hubbard model with different bandwidths ($U=20$, $J=1$) as a function of the triplon density and hopping parameter $t^\text{hop}_2$ of the narrow band ($t^\text{hop}_1 =1$). AFM order (an extension of the equilibrium phase) appears for small triplon density, whereas a FM state is stabilized for intermediate photo-doping and low $t^\text{hop}_2$. For higher photo-doping a spin-singlet $\eta$-SC order is stabilized for smaller values of $t^\text{hop}_2$ and spin-triplet $\eta$-SC appears for very strong photo-doping when $t^\text{hop}_2 \sim t^\text{hop}_1$. (b) FM order parameter as a function of Hund coupling $J$ at triplon density $0.27$ for $t^\text{hop}_2=0.3$. The vanishing magnetization at small $J$ shows the role of Hund coupling in stabilizing the FM state. (c) Magnetization and $\eta^{x}_{oy}$ order in band $1$ and band $2$ for $t^\text{hop}_2=0.3$. The magnetization is dominated by band $2$, while the $\eta$-SC order is hosted by band $1$. The effective temperature in all calculations is kept at $\beta_\text{eff} \sim 55$.
}
\label{fig2}
\end{figure}

{\it{Photo-induced hidden orders}}: We first study the nonequilibrium phase diagram as a function of photo-doping (triplon density) and the hopping parameter of the narrow band ($t^\text{hop}_{2}$), see Fig.~\ref{fig2}. To probe the ferromagnetic (FM) and spin-singlet $\eta$-SC states, we add the seed terms $H_\text{seed} = B_\text{seed}\sum_{i}m_{i}+P_\text{seed}\sum_{i}(c^{\dagger}_{i,1\uparrow} c^{\dagger}_{i,1\downarrow} + c^{\dagger}_{i,2\uparrow} c^{\dagger}_{i,2\downarrow} + h.c. )$ to our Hamiltonian \eqref{eq_1}, where $B_\text{seed}$ and $P_\text{seed}$ are, respectively, small magnetic and superconducting seed fields. Due to the partial thermalization of the electrons in each Hubbard band, the photodoped state can be represented by an equilibrium state with fixed number of multiplets (triplons, doublons and singlons), described by the effective low energy model  \cite{SWT,eta_spintrip}
\begin{eqnarray}
H_\text{eff} &=& H_\text{hop}(P_{in}P_{jn-1}) + H_{s/o}(P_{in}P_{jn}) 
+ H_{\eta}(P_{in}P_{jn-2}),\nonumber\\
\label{eq_2}
\end{eqnarray}
where $P_{in}$ is the projection operator at site $i$ on states with particle number $n$. The first term in Eq.~\eqref{eq_2} is the hopping term $-\sum_{\left<ij\right>n}\sum_{\alpha\sigma} t^\text{hop}_{\alpha}c^{\dagger}_{i,\alpha\sigma} c_{j,\alpha\sigma}$ which acts on sites which differ in particle number by one. The second term is the spin-orbital term  $2(t^{\text{hop}}_{\alpha})^2/U \sum_{\left<ij\right>} \left( \frac{1}{2} + 2\boldsymbol{s_i} . \boldsymbol{s_j} \right) \left( \frac{1}{2} + 2\boldsymbol{\tau_i} . \boldsymbol{\tau_j} \right)$ and acts on sites with the same particle numbers, where $\boldsymbol{s_i}$ and $\boldsymbol{\tau_i}$ are spin and orbital pseudospin operators at site $i$. Hund coupling $J>0$ ($J<0$) favors spin (orbital) order. The last term is the $\eta$-pseudospin term given by $-4(t^{\text{hop}}_{\alpha})^2/U \sum_{\left<ij\right>,\nu} \left( \eta_{i\nu}^{+}\eta_{j\nu}^{-} + \eta_{i\nu}^{-}\eta_{j\nu}^{+} \right)$ with $\nu$ an index for the six different SC orders (three spin-singlet orbital-triplet and three  spin-triplet orbital-singlet orders, see SM). This term acts on sites which differ in particle number by two \cite{eta_spintrip}.

Near zero photo-doping, there are negligible charge excitations and the system is governed by the $H_{s/o}$ term, which stabilizes the antiferromagnetic (AFM) phase with a transition temperature ($T_c$) of the order of $(t^\text{hop}_{\alpha})^{2}/U$. As the photo-doping is increased, the diffusing charge carriers melt the AFM phase, which results in a nonthermal PM metal. As the bandwidth of the narrower band ($t^\text{hop}_2$) is decreased, the $T_c$ also decreases, since we are on the strongly correlated side of the AFM dome. This shifts the boundary between the AFM and PM phases towards lower photo-doping. At intermediate photo-doping, the hopping term $H_\text{hop}$ starts to dominate, because of the substantial density of singly occupied (singlon), doubly occupied (doublon) and triply occupied (triplon) sites. If we decrease $t^\text{hop}_2$ in this regime to localize the electrons in the narrower band, a nonthermal FM state is stabilized by the interplay of Hund coupling and kinetic energy. A positive Hund coupling $J$ favors high-spin doublon states. On the other hand, the hopping processes in the wide band from doublon to singlon and from triplon to doublon sites, illustrated in the bottom panel of Fig.~\ref{fig1}, stabilize the FM state since this state allows to gain kinetic energy while keeping the local spin arrangement unchanged. Any other hopping processes would create a configuration with higher energy. Thus a nonthermal FM phase appears at lower $t^\text{hop}_2$ values and near triplon/singlon density 0.25 for $J>0$. In the absence of Hund coupling, all the local doublon states are degenerate in energy and the hopping term is not expected to favor the FM state. Indeed, by decreasing the Hund coupling, the FM magnetization disappears below a critical value of $J$, as shown in Fig.~\ref{fig2}(b).

\begin{figure}[t]
\includegraphics[width=0.4\textwidth]{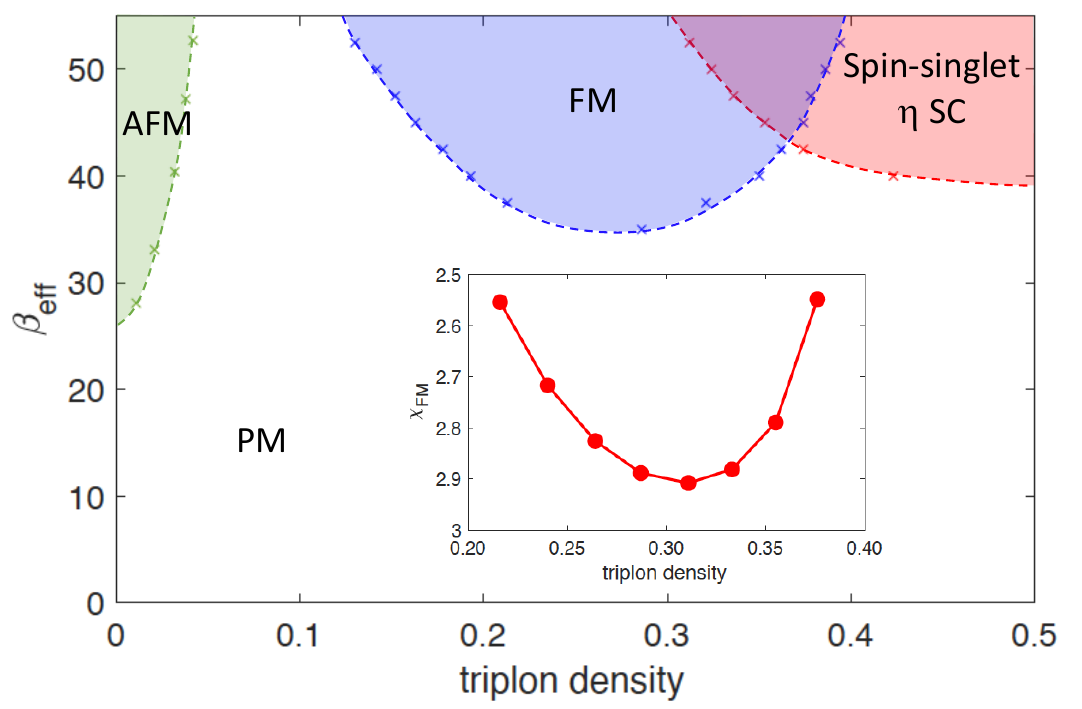}
\caption{NESS phase diagram of the two-orbital Hubbard model with $U=20$, $J=1$, $t^\text{hop}_1=1$ and $t^\text{hop}_2=0.4$ in the space of $\beta_\text{eff}$ and triplon density. The AFM phase appears in the undoped and weakly photo-doped region, while the $\eta$ SC state is stabilized in the high photo-doping region. At intermediate photo-doping we have the FM state. In these simulations, $T_b$ is varied to achieve different $\beta_\text{eff}$. The inset of the figure shows the FM susceptibility $\chi_\text{FM}$, estimated from the real-time calculations, as a function of doping. The optimal doping condition for FM order is consistent in both calculations.
}
\label{fig3}
\end{figure}

When the photo-doping is increased further, more charge excitations are created and for $t^\text{hop}_1 = t^\text{hop}_2 = 1$, in the high doping limit (triplon density $\gtrsim 0.46$) spin-triplet orbital-singlet $\eta$-SC order ($\eta^{x}_{sy}$) develops \cite{eta_spintrip}. As shown in Fig.~\ref{fig2}(a), this order disappears when $t^\text{hop}_2$ is decreased. For small $t^\text{hop}_2$, the charge excitations are mostly concentrated in the wider band, while the narrower band remains close to half-filled (e.~g., for $t^\text{hop}_1=1$, $t^\text{hop}_2 =0.3$, triplon density $\sim 0.28$ and $\beta_\text{eff} \sim 55$, the double occupancies in the wider band and the narrower band are 0.25 and 0.03, respectively). At sufficiently low effective temperature, the doublons and singlons in the wide band condense to give rise to spin-singlet orbital-triplet $\eta_{oy}$ order. Since the dominant contribution to the $\eta_{oy}$ order comes from the wider band ($c^{\dagger}_{i,1\uparrow} c^{\dagger}_{i,1\downarrow}$), we have an orbital selective $\eta_{oy}$ order (see Fig.~\ref{fig2}(c)).

Interestingly, in the photo-doping region with triplon density between $0.3$ to $0.4$, both the first and last term in Eq.~\eqref{eq_2} contribute and we find a coexistence of FM and $\eta$-SC order. At first sight, this is surprising, since magnetic order is usually detrimental to singlet pairing. Here, such a co-existence is possible because the two photo-induced order parameters are supported by different orbitals -- the wider band hosts the SC order, while the FM order is dominated by the spin-aligned localized electrons in the narrower band, as shown in Fig.~\ref{fig2}(c).

Next, we fix the $t^\text{hop}_1 =1, t^\text{hop}_2 = 0.4$ and investigate the phase diagram in the space of triplon density and inverse effective temperature $\beta_\text{eff}$, see Fig.~\ref{fig3}. At low $T_\text{eff}$ (high $\beta_\text{eff} \sim 50$), we have AFM order at low photo-doping. Around triplon density 0.25 the FM phase appears, while at large photo-doping, we find the $\eta_{oy}$-SC phase, which partly  coexists with the FM phase. With increasing temperature (decreasing $\beta_\text{eff}$) the ordered regions shrink and they disappear at similar $T_c$ values. The highest effective $T_{c}$ for the nonthermal FM order is around triplon density $\sim 0.29$, which is consistent with the optimal doping deduced from the phase diagram in Fig.~\ref{fig2}(a). The $\eta_{oy}$ phase shows a similar temperature dependence as in the one band Hubbard model \cite{eta_jiajun}, consistent with our observation that the $\eta_{oy}$-SC state is essentially decoupled from the narrow band and has a single-band characteristics. 

We see from Fig.~\ref{fig3} that both the nonthermal FM and $\eta$-SC orders appear at $\beta_\text{eff} > 30$. While this is a high temperature for realistic parameters, it is difficult to realize these effective temperatures in real-time simulations of a photoexcited multi-orbital system, even using the entropy cooling technique \cite{entropy_cooling}. However, near the optimal photo-doping density, we still expect to find an enhanced FM susceptibility at the accessible higher $T_\text{eff}$. 

\begin{figure}[t]
\includegraphics[width=0.4\textwidth]{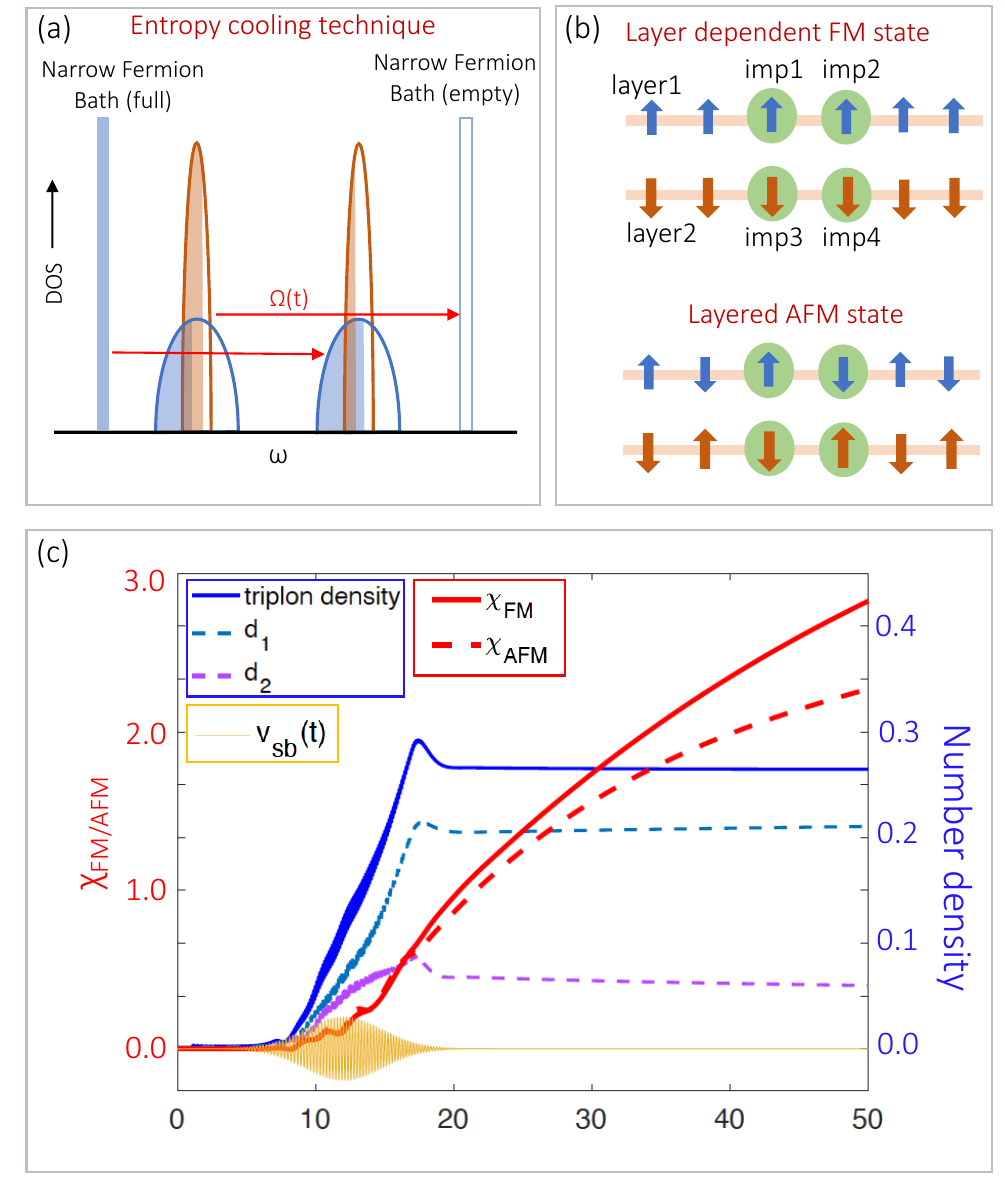}
\caption{(a) Illustration of the entropy cooling technique, where electrons are injected into the upper Hubbard bands from a fully occupied narrow Fermion bath and electrons are ejected from the lower Hubbard bands to an empty Fermion bath. (b) Schematic diagram of the four-impurity DMFT calculation mimicking the dynamics in two bilayer systems -- an AFM system (bottom) and an AFM stacked FM system (top). (c) Magnetic susceptibility $\chi_\text{FM/AFM}$ in the FM and AFM state (red curves) and time-evolution of various local observables -- triplon density, double occupancy in orbital $1$ ($d_1$) and orbital $2$ ($d_2$). The yellow curve shows the hopping pulse $v_{sb}(t)$ (not to scale) used to produce the photo-doping.
}
\label{fig5}
\end{figure}

{\it{Real-time simulations}}: To investigate the dynamics and measure the magnetic suceptibilities, we employ the entropy cooling technique \cite{entropy_cooling, entropy_cooling1} to minimize the heating effect in nonequilibrium photo-doped Mott insulators.  In this setup, we start with a half-filled Mott insulating state and add two additional narrow noninteracting Fermion baths (one completely full and one completely empty) as shown in Fig.~\ref{fig5}(a). These baths are coupled to the Mott insulating system by a hopping term $v_{sb}$. To facilitate an efficient photo-doping, while keeping the effective temperature of the system low, the hopping term is modulated in time as $v_{sb}(t)=A_{0}f_{\text{en}}(t)\sin(\Omega(t)t)$, where $A_{0}=1$ denotes the amplitude of the hopping, $f_{\text{en}}(t)$ is an envelope function determining the switch-on and switch-off times of the hopping and $\Omega(t)$ is a time-dependent (chirped) frequency. In our calculations with $U=20$, $J=1$ model, we chose $f_{\text{en}}(t)=\exp(-((t-t_{0})/2\sigma)^2)$ as a Gaussian with $t_0=12$ and $\sigma=2.8$. $\Omega(t)$ is increased linearly from an initial value $\Omega_{\text{ini}}=27.0$ to a final value $\Omega_{\text{fin}}=29.8$. The hopping modulation pulse
is shown in Fig.~\ref{fig5}(c) (yellow curve). In this set-up, the DMFT self-consistency condition becomes
\begin{eqnarray}
    \Delta(t,t^{\prime})=t_{0}(t)^{2}\gamma G(t,t^{\prime})\gamma + \sum_{\alpha}v_{sb}(t)^{2} G_{\text{bath},\alpha}(t,t^{\prime}) \hspace{5mm}
\end{eqnarray}
where $t_{0}(t)=t_{0}$ is kept fixed, $G(t,t^{\prime})$ is the Green's function of the system and $G_{\text{bath},\alpha}(t,t^{\prime})$ is the Green's function of the noninteracting bath with flavor $\alpha$ (full,empty).

Because of the conservation of total spin in our Hamiltonian, it is not possible to simulate the switching to a purely FM state within this framework. Thus, in order to study the photo-induced FM state, we implemented a four-impurity calculation, which mimics the bi-layer system shown in Fig.~\ref{fig5}(b). Here, the impurity models $1$ and $2$ represent two neighboring sites in layer 1, whereas the impurity models $3$ and $4$ represent the corresponding sites in layer 2. In this set-up, an AFM state is identified by AFM spin alignments in all four pairs of neighboring impurities, while a FM state is indicated by a layer-dependent magnetization, e.~g. with spin-up polarization on impurities $1$ and $2$ and spin-down polarization on impurities $3$ and $4$ (see Fig.~\ref{fig5}(b)).

In Fig.~\ref{fig5}(c) we show the result of the real-time DMFT calculation. As soon as the coupling to the narrow baths is switched on, the system gets photo-doped, as indicated by the increasing triplon density during the pulse. As a result of the different bandwidths, the double occupancy in orbital 1 ($d_1$) grows larger than in orbital 2 ($d_2$), which is consistent with the results of the NESS calculations. The magnetic susceptibility is measured by dividing the magnetization (magnetic order parameter) by the corresponding seed field as $\chi = \left< m \right>/B_\text{seed}$. In Fig.~\ref{fig5}(c), we show the $\chi_\text{FM}$ and $\chi_\text{AFM}$ estimates for FM and AFM order. In the FM case, we choose $B_\text{seed}>0$ for impurity 1 and 2, $B_\text{seed}<0$ for impurity 3 and 4, while in the AFM case $B_\text{seed}>0$ for impurity 1 and 4 and $B_\text{seed}<0$ for impurity 2 and 3. The FM susceptibility grows larger than the AFM one and keeps increasing at the longest simulation times, indicating a dominant tendency towards FM ordering. We can also control the photo-doping concentration by controlling the final frequency $\Omega_{\text{fin}}$ of the chirped pulse and extract $\chi_\text{FM}$ as a function of the triplon density, as shown in the inset of Fig.~\ref{fig3}. We find that $\chi_\text{FM}$ reaches a maximum value near triplon density $\sim 0.3$, which agrees with the optimal photodoping concentration for the FM phase in the NESS simulations. Due to numerical limitations, we cannot drive the system slow enough to reduce $T_\text{eff}$ further or drive the system for long enough time to demonstrate an actual symmetry breaking. But the close correlation between the FM phase boundary in the NESS approach and the FM susceptibility in the real-time simulations supports the appearance of a nonthermal FM phase in photo-doped two-orbital Mott insulators.

In conclusion, we demonstrated that the photo-doped two-band Hubbard model with orbital-dependent bandwidths hosts nonthermal FM and staggered SC order, and even a co-existence of both orders. Manganites, which are often described by the double exchange model, and NiO, with two Mott insulating half-filled $e_g$ bands, would be possible candidates to search for these nonthermal electronic orders.

{\it Acknowledgements}: We thank Y. Murakami for helpful comments on the manuscript and the Swiss National Science Foundation for the funding (Grant No.~200021-196966). The nonequilibrium DMFT calculations are based on the NESSi library \cite{NESSi} and the NESS simulations on a code originally developed by J.~Li and M.~Eckstein. The calculations were run on the beo06 cluster at the University of Fribourg.


\begin{thebibliography}{99}

\bibitem{lat_struc} H. Ichikawa, S. Nozawa, T. Sato, A. Tomita, K. Ichiyanagi, M. Chollet, L. Guerin, N. Dean, A. Cavalleri, S.-i. Adachi, {\it et al.}, Nat. Mater. {\bf 10}, 101 (2011).

\bibitem{mang_hid} L. Stojchevska, I. Vaskivskyi, T. Mertelj, P. Kusar, D. Svetin, S. Brazovskii, and D. Mihailovic, Science {\bf 344}, 177 (2014).

\bibitem{Fausti2011} D. Fausti, R. I. Tobey, N. Dean, S. Kaiser, A. Dienst, M. C. Hoffmann, S. Pyon, T. Takayama, H. Takagi, and A. Cavalleri, Science {\bf 331}, 189 (2011).

\bibitem{Cavalleri2018} A. Cavalleri, Contemp. Phys. {\bf 59}, 31 (2018).

\bibitem{Torre2021} A. de la Torre, D. M. Kennes, M. Claassen, S. Gerber, J. W. McIver, and M. A. Sentef, Rev. Mod. Phys. {\bf 93}, 041002 (2021).

\bibitem{ph_noneq} Y. Murakami, D. Golež, M. Eckstein, P. Werner, arXiv:2310.05201 (2023).

\bibitem{Dagotto2005} E. Dagotto, Science 309, 257 (2005). 

\bibitem{yuta_etasc} Y. Murakami, S. Takayoshi, T. Kaneko, A. M. Läuchli, and P. Werner, Phys. Rev. Lett. {\bf 130}, 106501 (2023).

\bibitem{yuta_etasc1} Y. Murakami, S. Takayoshi, T. Kaneko, Z. Sun, D. Golež, A. J. Millis, and P. Werner, Comm. Phys. {\bf 5}, 23 (2022).

\bibitem{opt_magord} A. Kirilyuk, A. V. Kimel, and T. Rasing, Rev. Mod. Phys. {\bf 82}, 2731 (2010).

\bibitem{ind_sc_cuprate} D. Fausti, R. I. Tobey, N. Dean, S. Kaiser, A. Dienst, M. C. Hoffmann, S. Pyon, T. Takayama, H. Takagi, and A. Cavalleri, Science {\bf 331}, 189 (2011).

\bibitem{phind_sc} M. Mitrano, A. Cantaluppi, D. Nicoletti, S. Kaiser, A. Perucchi, S. Lupi, P. Di Pietro, D. Pontiroli, M. Ricco, S. R. Clark, et al., Nature {\bf 530}, 461 (2016).

\bibitem{exciton_insulator} P. Werner, and Y. Murakami, Phys. Rev. B {\bf 102}, 241103(R) (2020).

\bibitem{opt_exint} R. Mikhaylovskiy, E. Hendry, A. Secchi, J. H. Mentink, M. Eckstein, A. Wu, R. Pisarev, V. Kruglyak, M. Katsnel-son, T. Rasing, et al., Nat. Commun. {\bf 6}, 8190 (2015).

\bibitem{tri_chiral} J. Li, M. M\"uller, A. Kim, A. L\"aeuchli, and P. Werner, Phys. Rev. B {\bf 107}, 205115 (2023).

\bibitem{Kugel_Khomskii} J. Li, H. U. R. Strand, P. Werner and M. Eckstein , Nat. Com. {\bf 9}, 4581 (2018).

\bibitem{eta_jiajun} J. Li, D. Golez, P. Werner, and M. Eckstein, Phys. Rev. B {\bf 102}, 165136 (2020)

\bibitem{eta_spintrip} S. Ray, Y. Murakami, and P. Werner, Phys. Rev. B {\bf 108}, 174515 (2023).

\bibitem{kaneko1} S. Ejima, T. Kaneko, F. Lange, S. Yunoki, and H. Fehske, Phys. Rev. Research {\bf 2}, 032008(R) (2020).

\bibitem{kaneko2} T. Kaneko, S. Yunoki, and A. J. Millis, Phys. Rev. Research {\bf 2}, 032027(R) (2020).

\bibitem{Sensarma2010} R. Sensarma, D. Pekker, E. Altman, E. Demler, N. Strohmaier, D. Greif, R. J\"ordens, L. Tarruell, H. Moritz, and T. Esslinger, Phys. Rev. B {\bf 82}, 224302 (2010).

\bibitem{Eckstein2011} M. Eckstein and P. Werner, Phys. Rev. B {\bf 84}, 035122 (2011).

\bibitem{Strohmaier2010} N. Strohmaier, D. Greif, R. Jördens, L. Tarruell, H. Moritz, T. Esslinger, R. Sensarma, D. Pekker, E. Alt- man, and E. Demler, Phys. Rev. Lett. {\bf 104}, 080401 (2010).

\bibitem{Lenarcic2014} Z. Lenarčič and P. Prelovšek, Phys. Rev. B {\bf 90}, 235136 (2014). 

\bibitem{VO2_dynamics} A. Cavalleri, C. Tóth, C. W. Siders, J. A. Squier, F. Ráksi, P. Forget, and J. C. Kieffer, Phys. Rev. Lett. {\bf 87}, 237401 (2001).

\bibitem{ferro_chrgtrnsfr_dynamics} E. Collet, M. H. Lemée-Cailleau, M. Le Cointe, H. Cailleau, M. Wulff, T. Luty, S. Koshihara, M. Meyer, L. Toupet, P. Rabiller, and S. Techert, Science {\bf 300}, 612 (2003).

\bibitem{mag_switch} T. Li, A. Patz, L. Mouchliadis, J. Yan, T. A. Lograsso, I. E. Perakis, and J. Wang, Nature {\bf 496}, 69 (2013).

\bibitem{Qint_Mott} S. Wall, D. Brida, S. R. Clark, H. P. Ehrke, D. Jaksch, A. Ardavan, S. Bonora, H. Uemura, Y. Takahashi, T. Hasegawa, H. Okamoto, G. Cerullo, and A. Cavalleri Nature Phys. {\bf 7}, 114 (2011).

\bibitem{opt_magcont} J. H. Mentink, J. Phys.: Condens. Matter {\bf 29}, 453001 (2017)

\bibitem{Ishihara_2007} H. Matsueda and S. Ishihara, J. Phys. Soc. Jpn. {\bf 76}, 083703 (2007).

\bibitem{Ishihara_2011} Y. Kanamori, H. Matsueda, and S. Ishihara, PRL {\bf 107}, 167403 (2011)

\bibitem{Ishihara_DE1} Y. Kanamori, H. Matsueda, and S. Ishihara, Phys. Rev. Lett. {\bf 103}, 267401 (2009).

\bibitem{Ishihara_DE2} Y. Kanamori, H. Matsueda, and S. Ishihara, Phys. Rev. B {\bf 82}, 115101 (2010).

\bibitem{manganites_1} K. Miyano, T. Tanaka, Y. Tomioka, and Y. Tokura, Phys. Rev. Lett. {\bf 78}, 4257 (1997).

\bibitem{Qdyn_spin_struc} W. Koshibae, N. Furukawa, and N. Nagaosa, Phys. Rev. Lett. {\bf 103}, 266402 (2009).

\bibitem{ultrafast_demag1} E. Beaurepaire, J. C. Merle, A. Daunois, and J. Y. Bigot, Phys. Rev. Lett. {\bf 76}, 4250 (1996).

\bibitem{ultrafast_demag2} B. Koopmans, G. Malinowski, F. Dalla Longa, D. Steiauf, M. Fähnle, T. Roth, M. Cinchetti, and M. Aeschlimann, Nat. Mater. {\bf 9}, 259 (2010).

\bibitem{manganites_2} M. Fiebig,  K. Miyano, Y. Tomioka, Y. Tokura, Science 280, 1925 (1998)

\bibitem{manganites_3} R. D. Averitt, A. I. Lobad, C. Kwon, S. A. Trugman, V. K. Thorsmølle, and A. J. Taylor, Phys. Rev. Lett. {\bf 87}, 017401 (2001).

\bibitem{Ishihara_DE3} A. Ono, S. Ishihara, Phys. Rev. B {\bf 98}, 214408 (2018).

\bibitem{NESS_martin} J. Li and M. Eckstein, Phys. Rev. B {\bf 103}, 045133 (2021).

\bibitem{Keiter_1970} H. Keiter and J. C. Kimball, Phys. Rev. Lett. {\bf 25}, 672 (1970).

\bibitem{Eckstein_2010} M. Eckstein and P. Werner, Phys. Rev. B {\bf 82}, 115115 (2010).

\bibitem{SWT} S.-S. B. Lee, J. v. Delft, and A. Weichselbaum, Phys. Rev. B {\bf 96}, 245106 (2017).

\bibitem{entropy_cooling} P. Werner, M. Eckstein, M. Müller, and G. Refael, Nat. Comm. {\bf 10}, 5556 (2019).

\bibitem{entropy_cooling1} P. Werner, J. Li, D. Golez, and M. Eckstein, Phys. Rev. B {\bf 100}, 155130 (2019).

\bibitem{NESSi} M. Schüler, D. Golež, Y. Murakami, N. Bittner, A. Herrmann, H. U. R. Strand, P. Werner, and M. Eckstein, Comput. Phys. Commun. {\bf 257}, 107484 (2020).

\end{thebibliography}
\end{document}


\title{Photo-induced Ferromagnetic and Superconducting Orders in Multi-orbital Hubbard Models \\ Supplementary Material}
\author{Sujay Ray}
\affiliation{Department of Physics, University of Fribourg, 1700 Fribourg, Switzerland}
\author{Philipp Werner}
\affiliation{Department of Physics, University of Fribourg, 1700 Fribourg, Switzerland}
\date{\today}
\maketitle

\section{Effective low energy model}

This section provides additional information on the effective low energy model which is used in the main text. We start with the two-orbital Hubbard Hamiltonian with orbital dependent hopping parameters, as defined in Eq.~(1) of the main text. To define an effective low energy Hamiltonian, we introduce projected particle creation operators $\tilde{c}^{\dagger}_{i,\alpha\sigma,n}=P_{in}c^{\dagger}_{i,\alpha\sigma}$ and annihilation operators $\tilde{c}_{i,\alpha\sigma,n}=P_{in}c_{i,\alpha\sigma}$, where $P_{in}$ is the projector to states at site $i$ with $n$ particles \cite{SWT, eta_spintrip}. In our two-orbital model, each site can host at most $4$ particles (two spin degrees of freedom and two orbital degrees of freedom), which means that $n$ can take values from $0$ to $4$. Using this notation, we can perform a rotating frame transformation to arrive at the effective low energy Hamiltonian (neglecting the three site terms) given by \cite{SWT, eta_spintrip}
\begin{eqnarray}
H_\text{eff} &=& H_{\mu} + H_{\text{hop}} + H_{(t^{\text{hop}})^2/U} + O((t^{\text{hop}})^3/U),\label{eq_heff}
\end{eqnarray}
where,
\begin{eqnarray}
H_{\text{hop}} &=& -\sum_{\left<ij\right>n}\sum_{\alpha\sigma} t^\text{hop}_{\alpha}c^{\dagger}_{i,\alpha\sigma} c_{j,\alpha\sigma} P_{in-1}P_{jn} + (i \leftrightarrow j),\\
H_{(t^{\text{hop}})^2/U} &=& \sum_{\left<ij\right>} \sum_{n} \left( \sum_{\substack{\alpha,\alpha^{\prime} \\ \sigma,\sigma^{\prime}}} \sum_{\alpha\sigma\ne\alpha^{\prime}\sigma^{\prime}} \frac{t^\text{hop}_{\alpha}t^\text{hop}_{\alpha^{\prime}}}{U} c^{\dagger}_{i,\alpha\sigma} c^{\dagger}_{i,\alpha^{\prime}\sigma^{\prime}} c_{j,\alpha^{\prime}\sigma^{\prime}} c_{j,\alpha\sigma} \right) P_{in-2}P_{jn} \nonumber \\
&& \hspace{0cm} + \sum_{\left<ij\right>} \sum_{n, m} \frac{t^\text{hop}_{\alpha}t^\text{hop}_{\alpha^{\prime}}}{mU} \left( n_i - \sum_{\substack{\alpha,\alpha^{\prime} \\ \sigma,\sigma^{\prime}}}c^{\dagger}_{i,\alpha\sigma} c_{i,\alpha^{\prime}\sigma^{\prime}} c^{\dagger}_{j,\alpha^{\prime}\sigma^{\prime}} c_{j,\alpha\sigma} \right) P_{in}P_{jn-m-1} + (i \leftrightarrow j).
\label{eq_eff}
\end{eqnarray}
In the above expressions, we can identify the three terms mentioned in Eq. (4) of the main text. $H_{\text{hop}}$ is the hopping term, which acts on neighboring sites $j$ with particle number $n$ and $i$ with particle number $n-1$, or vice versa. Similarly, the first term of $H_{(t^{\text{hop}})^2/U}$ acts on sites $j$ and $i$ with particle numbers $n-2$ and $n$. This term can be expressed in terms of six $\eta$-SC operators \cite{eta_spintrip, Hoshino_spinfreezing} -- three spin-singlet orbital-triplet operators, 
\begin{eqnarray}
\eta^{+}_{i,ox}= \frac{1}{2} \delta^{A/B}_{i} \left( c^{\dagger}_{i,1\uparrow} c^{\dagger}_{i,1\downarrow} - c^{\dagger}_{i,2\uparrow} c^{\dagger}_{i,2\downarrow} \right), \nonumber \\
\eta^{+}_{i,oy}= \frac{1}{2} \delta^{A/B}_{i} \left( c^{\dagger}_{i,1\uparrow} c^{\dagger}_{i,1\downarrow} + c^{\dagger}_{i,2\uparrow} c^{\dagger}_{i,2\downarrow} \right), \nonumber \\
\eta^{+}_{i,oz}= \frac{1}{2} \delta^{A/B}_{i} \left( c^{\dagger}_{i,1\uparrow} c^{\dagger}_{i,2\downarrow} - c^{\dagger}_{i,1\uparrow} c^{\dagger}_{i,2\downarrow} \right),
\end{eqnarray}
and three spin-triplet orbital-singlet operators, 
\begin{eqnarray}
\eta^{+}_{i,sx}= \frac{1}{2} \delta^{A/B}_{i} \left( c^{\dagger}_{i,1\uparrow} c^{\dagger}_{i,2\uparrow} - c^{\dagger}_{i,1\downarrow} c^{\dagger}_{i,2\downarrow} \right), \nonumber \\
\eta^{+}_{i,sy}= \frac{1}{2} \delta^{A/B}_{i} \left( c^{\dagger}_{i,1\uparrow} c^{\dagger}_{i,2\uparrow} + c^{\dagger}_{i,1\downarrow} c^{\dagger}_{i,2\downarrow} \right), \nonumber \\
\eta^{+}_{i,sz}= \frac{1}{2} \delta^{A/B}_{i} \left( c^{\dagger}_{i,1\uparrow} c^{\dagger}_{i,2\downarrow} - c^{\dagger}_{i,1\uparrow} c^{\dagger}_{i,2\downarrow} \right),
\end{eqnarray}
where $\delta^{A/B}_{i}$ takes the values $+1$ $(-1)$ on sublattice A (B). With the help of these $\eta$ operators, we can write the first term of $H_{t_{\text{hop}}^2/U}$ as
\begin{eqnarray}
H_{\eta}=O\left[-4(t^{\text{hop}}_{\alpha})^2/U\right] \sum_{\left<ij\right>,\nu} \left( \eta_{i\nu}^{+}\eta_{j\nu}^{-} + \eta_{i\nu}^{-}\eta_{j\nu}^{+} \right),
\end{eqnarray}
where $\nu=(ox,oy,oz,sx,sy,sz)$ and $\eta^{-}_{i,\nu}$ is the Hermitian conjugate of $\eta^{+}_{i,\nu}$. The coefficient $O[-4(t^{\text{hop}}_{\alpha})^2/U]$ depends on the type of $\eta$ pairing, e,g,. for $c^{\dagger}_{i,1\uparrow} c^{\dagger}_{i,1\downarrow}$, we have $-4(t^{\text{hop}}_{1})^2/U$ and for $c^{\dagger}_{i,1\uparrow} c^{\dagger}_{i,2\downarrow}$, we have $-4(t^{\text{hop}}_{1}t^{\text{hop}}_{2})/U$. This also shows that for different bandwidths $t^{\text{hop}}_{1} > t^{\text{hop}}_{2}$, $\eta$ pairing in band $1$ is favored because of the higher value of this coefficient. We also define $\eta^{x}_{i,\nu}=\text{Re}\left[ \eta^{+}_{i,\nu} \right]$, which is used in the main text. 

The second term in $H_{(t^{\text{hop}})^2/U}$ acts on sites $i$ and $j$ with particle numbers $n$ and $n-m-1$ ($m\ne 0$) and in general gives rise to a spin-orbital term 
\begin{eqnarray}
H_{s/o}=-\frac{4t^{\text{hop}}_{1}t^{\text{hop}}_{2}}{mU} \sum_{\left<ij\right>} \left( \frac{1}{2} + 2\boldsymbol{s_i} . \boldsymbol{s_j} \right) \left( \frac{1}{2} + 2\boldsymbol{\tau_i} , \boldsymbol{\tau_j} \right)P_{i,n}P_{j,n-m-1},
\end{eqnarray}
where $\boldsymbol{s}_i=\sum_{\alpha}c^{\dagger}_{i,\alpha\sigma}\boldsymbol{\sigma}_{\sigma\sigma^{\prime}}c_{i,\alpha\sigma^{\prime}}$ is the spin moment and $\boldsymbol{\tau}_i=\sum_{\sigma}c^{\dagger}_{i,\alpha\sigma}\boldsymbol{\sigma}_{\alpha\alpha^{\prime}}c_{i,\alpha^{\prime}\sigma}$ is the orbital pseudo-spin moment at site $i$ with $\boldsymbol{\sigma}$ denoting the Pauli spin matrices. Since in our nonequilibrium system we have singlons, doublons and triplons, we consider $m=-1,1$ and $m=-3$. For $m=1$ and $-3$,  the particle numbers on site $i$ and $j$ become $n$ and $n+2$, and since the coefficient $\frac{2}{3}\frac{4t^{\text{hop}}_{1}t^{\text{hop}}_{2}}{U}$ is smaller than the coefficient of $H_{\eta}$, this does not contribute considerably between a singlon site and a triplon site. For $m=-1$, the particle numbers on sites $i$ and $j$ are equal ($n$) and this gives rise to a Heisenberg exchange interaction at half filling with antiferromagnetic exchange coupling $\frac{4t^{\text{hop}}_{1}t^{\text{hop}}_{2}}{U}$. Thus we arrive at the effective low energy Hamiltonian presented in Eq. (4) of the main text.

\section{DMFT calculations}

\begin{figure}[b]\centering
\includegraphics[width=0.96\textwidth]{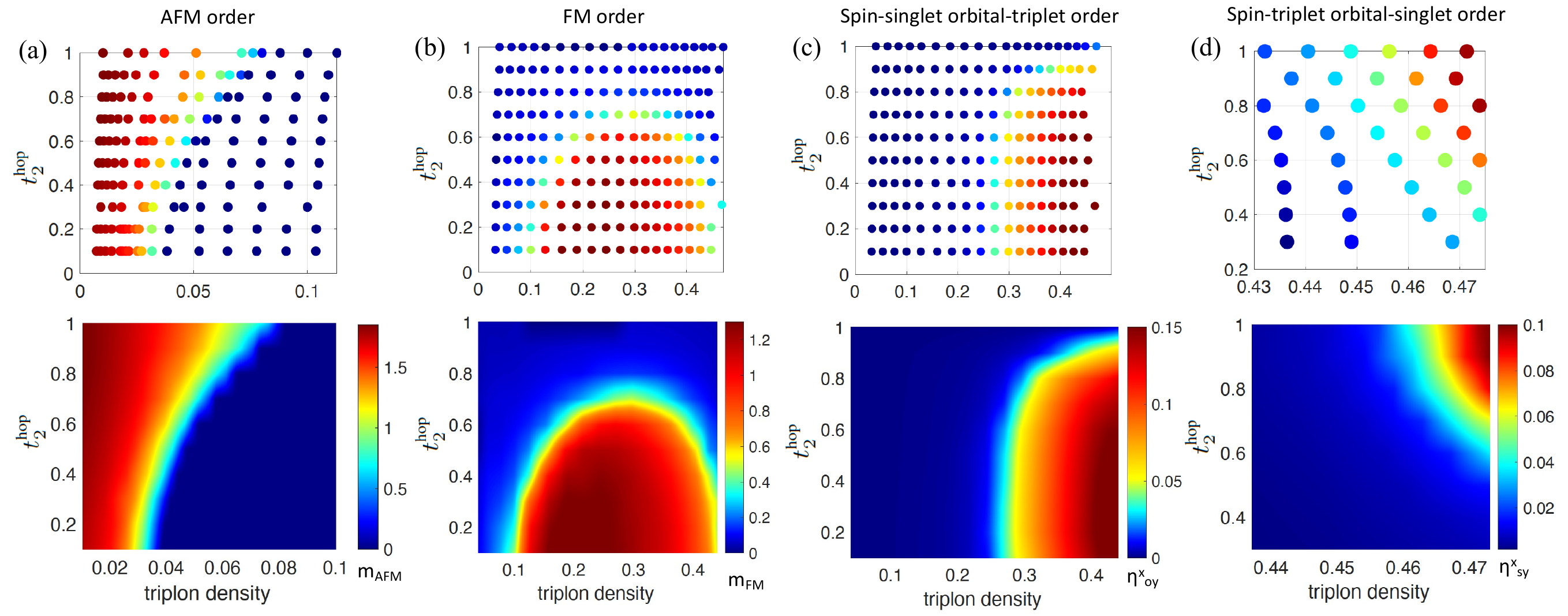}
\caption{Nonequilibrium DMFT data for (a) AFM, (b) FM, (c) $\eta^{x}_{oy}$ superconductivity, and (d) $\eta^{x}_{sy}$ superconductivity in the space of hopping parameter of the narrow band $t^\text{hop}_{2}$ and triplon density for $U=20, J=1$ at $\beta_\text{eff} \sim 55 $. The upper panel shows the DMFT data points. In the lower panel, interpolated results are plotted to better identify the phase boundaries.
}
\label{fig_SM1}
\end{figure}

As mentioned in the main text, we use the spinor basis $\psi^{\dagger}=(c_{1\uparrow}^{\dagger}, \hspace{0.2cm} c_{2\uparrow}, \hspace{0.2cm} c_{1\downarrow}^{\dagger}, \hspace{0.2cm} c_{2\downarrow})$ in order to measure spin-triplet superconducting order in the DMFT calculations. The Green's function $G(t,t^{\prime})$ and the hybridization function $\Delta(t,t^{\prime})$ in this Nambu basis become the $4\times4$ matrices
\begin{eqnarray}
\Delta(t,t^{\prime}) &=& \left[
\begin{array}{ c c c c }
\Delta_{1\uparrow 1\uparrow}^{c^{\dagger}c} & \Delta_{1\uparrow 2\uparrow}^{c^{\dagger}c^{\dagger}} & 0 & \Delta_{1\uparrow 2\downarrow}^{c^{\dagger}c^{\dagger}}\\
\Delta_{2\uparrow 1\uparrow}^{cc} & \Delta_{2\uparrow 2\uparrow}^{cc^{\dagger}} & \Delta_{2\uparrow 1\downarrow}^{cc} & 0 \\
 0 & \Delta_{1\downarrow 2\uparrow}^{c^{\dagger}c^{\dagger}} & \Delta_{1\downarrow 1\downarrow}^{c^{\dagger}c} & \Delta_{1\downarrow 2\downarrow}^{c^{\dagger}c^{\dagger}}\\
 \Delta_{2\downarrow 1\uparrow}^{cc} & 0 & \Delta_{2\downarrow 1\downarrow}^{cc} & \Delta_{2\downarrow 2\downarrow}^{cc^{\dagger}}\\
\end{array} \right], \ \
G(t,t^{\prime})=\left[
\begin{array}{ c c c c }
G_{1\uparrow 1\uparrow}^{cc^{\dagger}} & G_{1\uparrow 2\uparrow}^{cc} & 0 & G_{1\uparrow 2\downarrow}^{cc} \\
G_{2\uparrow 1\uparrow}^{c^{\dagger}c^{\dagger}} & G_{2\uparrow 2\uparrow}^{c^{\dagger}c} & G_{2\uparrow 1\downarrow}^{c^{\dagger}c^{\dagger}} & 0 \\
 0 & G_{1\downarrow 2\uparrow}^{cc} & G_{1\downarrow 1\downarrow}^{cc^{\dagger}} & G_{1\downarrow 2\downarrow}^{cc}\\
 G_{2\downarrow 1\uparrow}^{c^{\dagger}c^{\dagger}} & 0 & G_{2\downarrow 1\downarrow}^{c^{\dagger}c^{\dagger}} & G_{2\downarrow 2\downarrow}^{c^{\dagger}c}\\
\end{array} \right].
\end{eqnarray}
The DMFT self consistency equations in this Nambu basis can be written as $\Delta(t,t^{\prime})=t_{0}^2\gamma G(t,t^{\prime})\gamma$ . We can choose the $\gamma$ matrix to detect different orders. For example, $\gamma=$ diag( $1, \hspace{0.2cm} -1, \hspace{0.2cm} 1, \hspace{0.2cm} -1$ ) allows to detect uniform superconducting order, while for staggered $\eta$ superconducting order we choose $\gamma=$ diag( $1, \hspace{0.2cm} 1, \hspace{0.2cm} 1, \hspace{0.2cm} 1$ ). Similarly, we use the spinor basis $\psi^{\dagger}=(c_{1\uparrow}^{\dagger}, \hspace{0.2cm} c_{2\uparrow}^{\dagger}, \hspace{0.2cm} c_{1\downarrow}, \hspace{0.2cm} c_{2\downarrow})$ for measuring the spin-singlet superconducting order. The Green's function $G(t,t^{\prime})$ and the hybridization function $\Delta(t,t^{\prime})$ in this Nambu basis become
\begin{eqnarray}
\Delta(t,t^{\prime}) &=& \left[
\begin{array}{ c c c c }
\Delta_{1\uparrow 1\uparrow}^{c^{\dagger}c} & 0 & \Delta_{1\uparrow 1\downarrow}^{c^{\dagger}c^{\dagger}} & \Delta_{1\uparrow 2\downarrow}^{c^{\dagger}c^{\dagger}} \\
0 & \Delta_{2\uparrow 2\uparrow}^{c^{\dagger}c} & \Delta_{2\uparrow 1\downarrow}^{c^{\dagger}c^{\dagger}} & \Delta_{2\uparrow 2\downarrow}^{c^{\dagger}c^{\dagger}} \\
 \Delta_{1\downarrow 1\uparrow}^{cc} & \Delta_{1\downarrow 2\uparrow}^{cc} & \Delta_{1\downarrow 1\downarrow}^{cc^{\dagger}} & 0 \\
 \Delta_{2\downarrow 1\uparrow}^{cc} & \Delta_{2\downarrow 2\uparrow}^{cc} & 0 & \Delta_{2\downarrow 2\downarrow}^{cc^{\dagger}}\\
\end{array} \right], \ \
G(t,t^{\prime})=\left[
\begin{array}{ c c c c }
G_{1\uparrow 1\uparrow}^{cc^{\dagger}} & 0 & G_{1\uparrow 1\downarrow}^{cc} & G_{1\uparrow 2\downarrow}^{cc} \\
0 & G_{2\uparrow 2\uparrow}^{cc^{\dagger}} & G_{2\uparrow 1\downarrow}^{cc} & G_{2\uparrow 2\downarrow}^{cc} \\
 G_{1\downarrow 1\uparrow}^{c^{\dagger}c^{\dagger}} & G_{1\downarrow 2\uparrow}^{c^{\dagger}c^{\dagger}} & G_{1\downarrow 1\downarrow}^{c^{\dagger}c} & 0 \\
 G_{2\downarrow 1\uparrow}^{c^{\dagger}c^{\dagger}} & G_{2\downarrow 2\uparrow}^{c^{\dagger}c^{\dagger}} & 0 & G_{2\downarrow 2\downarrow}^{c^{\dagger}c}\\
\end{array} \right]. 
\end{eqnarray}
With the above diagonal $\gamma$ matrices, we can also simultaneously detect FM order. For AFM order, the $\gamma$ matrix is off-diagonal and given by
\begin{eqnarray}
\gamma^\text{AFM} = \left[
\begin{array}{ c c c c }
0 & 0 & 1 & 0 \\
0 & 0 & 0 & 1 \\
1 & 0 & 0 & 0 \\
0 & 1 & 0 & 0 \\
\end{array} \right]. 
\end{eqnarray}

\begin{figure}[t]\centering
\includegraphics[width=0.5\textwidth]{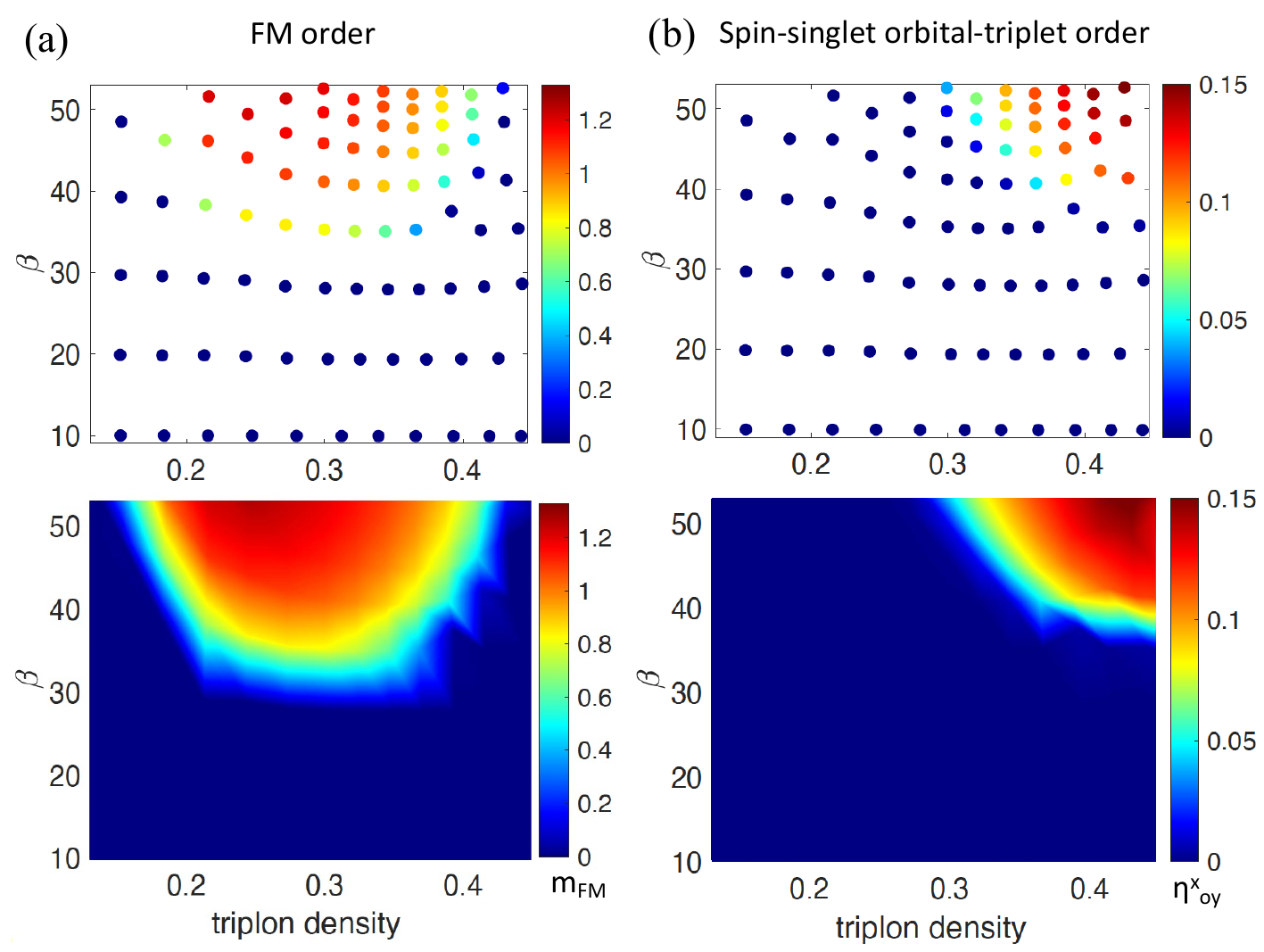}
\caption{Nonquilibrium DMFT data for (a) FM and (b) $\eta^{x}_{oy}$ superconductivity in the space of inverse effective temperature $\beta_{\text{eff}}$ and triplon density for $U=20, J=1$ and $t^{\text{hop}}_{2}=0.4$. The upper panel shows the DMFT data points. In the lower panel, interpolated results are shown to better identify the phase boundary.
}
\label{fig_SM2}
\end{figure}

To detect an ordered phase we apply a small seed field ($10^{-3}$) and couple it to the order parameter. We define the local magnetic moment as $m_i = n_{i,1\uparrow} + n_{i,2\uparrow} - n_{i,1\downarrow} - n_{i,2\downarrow}$. In a ferromagnetic (FM) state $m_\text{FM}=m_{i}$ has the same sign on each site, while in an antiferromagnetic (AFM) state $m_i$ has opposite signs on the A and B sublattices, so that $m_\text{AFM}=\delta^{A/B}_{i}m_{i}$. 

In Fig.~\ref{fig_SM1} we show the DMFT data used to map out the phase diagram for a photo-doped system with $U=20, J=1$ and $\beta_\text{eff}\sim 55$ in the space spanned by the hopping parameter $t^{\text{hop}}_{2}$ and the triplon density ($t^{\text{hop}}_{1}=1$). We show the values of all order parameters (AFM, FM, $\eta^{x}_{oy}$, $\eta^{x}_{sy}$) using a color map. The upper panels report the actual data points and the lower panels show the color maps of the interpolated data. From the interpolated results, we can extract the phase boundaries between the different phases. 

To change the triplon density, we change the chemical potential of the Fermion baths $\mu_b$. As shown in the phase diagram in the main text, the AFM phase is restricted to the low photodoping region, see also Fig.~\ref{fig_SM1}(a). The FM and $\eta^{x}_{oy}$ order are shown in Fig.~\ref{fig_SM1}(b) and (c). Spin-triplet orbital-singlet order $\eta^{x}_{sx}$ is concentrated in a small region of high photo-doping and $t^{\text{hop}}_{2} \sim 1$, as shown in Fig.~\ref{fig_SM1}(d). Similarly, in Fig.~\ref{fig_SM2}, we plot the DMFT data and the interpolated results for FM and $\eta^{x}_{oy}$-SC order as a function of triplon density and inverse effective temperature $\beta_{\text{eff}}$. Because of partial thermalization in each Hubbard band, the photodoped system can always be characterized by an effective temperature ($T_{\text{eff}}=1/\beta_{\text{eff}}$). In the NESS calculations, the effective temperature of the system is controlled by tuning the temperature of the Fermion baths $T_b$. We find that the FM order has the highest $T_c$ near half photo-doping, i.e. triplon density 0.25 (Fig.~\ref{fig_SM2}(a)), which is consistent with the optimal doping in Fig.~\ref{fig_SM1}(b). The spin-singlet $\eta^{x}_{oy}$ order has a similar temperature dependence as in the single band Hubbard model, which is consistent with our finding that $\eta^{x}_{oy}$ order is supported almost entirely by the wider band.